\documentclass[12pt]{article}
\usepackage{epsf}

\begin{document}

\begin{flushright}
SLAC--PUB--8794 \\
March 2001
\end{flushright}

\vfill

\begin{center}
{\bf\Large
     Standard Model Parameters and the Cosmological Constant\footnote{\baselineskip=12pt
Work supported by Department of Energy contract DE--AC03--76SF00515.}}

\bigskip

James D. Bjorken\\
Stanford Linear Accelerator Center\\
Stanford University, Stanford, CA 94309\\
\end{center}

\bigskip\bigskip

\begin{center}
{\large\bf  Abstract}
\end{center}

Simple functional relations amongst standard model couplings, including
gravitional, are conjectured.  Possible implications for cosmology and
future theory are discussed.
\vfill
\begin{center}
Submitted to The Physical Review.
\end{center}
\vfill
\newpage

\section{Introduction}

There are many oft-stated reasons for believing that the standard model
of elementary particles is incomplete.  One is the large number of
``fundamental" parameters, of order 20, for the ``old" standard model.
And this number increases to about 30 for the ``new" standard model,
which includes parameters describing neutrino masses and mixings.
Either way, one expects the future, better theory to contain fewer
parameters, implying that there should exist relationships between the
existing standard-model parameters.  The search for possible
relationships is the topic of this note.

It is most likely that such relationships are very complicated and
indirect.  Therefore the attempt to find them with the information at
hand can and should be viewed with deep suspicion and skepticism.  But
if there is a nonvanishing chance, however small, that simple,
discoverable relationships do exist, then it would seem that there is
little to lose by supposing that this is the case and engaging in the
pursuit.  It is this attitude, with eyes wide open, that we adopt here.

Existence of such relationships will be most likely if the amount of
``new physics" between accessible energies, at and below the electroweak
scale, and ultrahigh energies, at or beyond the grand-unification (GUT)
or Planck scale, is minimized \cite{ref:a}.  Therefore an implicit
assumption taken here is that new physics at energies between
electroweak and GUT scales is absent or of minimal importance.  This in
turn implies that the ``hierarchy problem", {\em i.e.} why the
quadratically divergent Higgs-boson mass is so small, is resolved at a
level deeper than supersymmetry at the electroweak scale, perhaps at the
same level as for the resolution of the ``cosmological constant
problem".  We shall return to this point in Section 6.

Existence of simple relationships between standard model parameters may
also imply that the dynamics of the future theory is relatively simple.
Otherwise, why should any such simple relationship exist?  This is an
additional stimulant for attempting the search.

Our considerations will proceed in three stages.  The starting point
will be a review of the standard model parameters and of the ``gaugeless
limit", which expresses in a way the conventional wisdom on how the
standard model is constructed.  We then discuss an intermediate version
which relates parameters in the gauge sector to those in the Higgs
sector.  Our final, most speculative step is to relate them all to
parameters in the gravitational sector.

\section{Standard Model Parameters}

The standard model parameters include the three gauge coupling
constants, which we here assume evolve from a common source at the GUT
scale characterized by a GUT fine-structure constant $\alpha_{\rm
gut}\approx 1/40$ or so.  Many of the remaining parameters lie in the
Higgs sector.  The two most important are the strength $v$ of the Higgs
condensate (the $vev$) and the strength $\lambda$ of the
elastic scattering amplitude of the Higgs boson with itself.  In
addition there are many Yukawa coupling constants $h_i$ of the Higgs
fields to quarks and leptons, responsible for their masses and mixings,
including CP violating effects.  By far the largest of these couplings
is that of the Higgs boson to the top quark, which we simply denote by
$h$.  In this note we lack the sophistication to consider the myriad of
smaller couplings, and will effectively set them to zero.

We shall keep both parameters from the gravitational sector.  The scale
of the Planck mass $M$, which determines Newton's constant, will be
considered equivalent to the GUT, grand-unification scale, because again
we will lack the sophistication to distinguish them.  However we shall
not neglect the cosmological-constant scale $\Lambda \sim \mu^4$, with
$\mu \sim 30\, cm^{-1}\sim 7 \times 10^{-4}\, eV$.

Finally, we shall have nothing to say about the parameter $\theta$,
which controls CP violation in the strong-interaction QCD sector, and
will set it to zero.

\section{The Gaugeless Limit}

At electroweak energy scales and above, all gauge couplings are small,
and it is a reasonable approximation, both for phenomenological and
conceptual purposes, to set them to zero, most efficiently by letting
$\alpha_{\rm gut} \rightarrow 0$.  The dynamics left behind is that of
the Higgs sector, which is brutally exposed in all of its intrinsic
ugliness.  The intermediate bosons become massless, allowing rapid decay
cascades of all quarks to the up quark.  All leptons, including the
electron, decay to neutrinos via (longitudinal) $W$ emission.  The
longitudinal $W$'s remain coupled in the gaugeless limit, emerging as
the massless Goldstone modes associated with spontaneous symmetry
breaking \cite{ref:b}.

This gaugeless limit quite accurately expresses (in reverse) the
textbook picture of how the standard-model electroweak dynamics works:
first the Higgs mechanism in isolation is constructed; then the effect
of the gauge interactions is included.  But it is possible
that this two-step approach is in the long run better viewed as a
linked, single step.  Something like this is expressed in the ``second
gaugeless limit", to which we now turn.

\section{The Second Gaugeless Limit}

If standard-model parameters are linked, it should be the case that if
$\alpha_{\rm gut}$ is set to zero, other standard-model parameters are
changed.  In searching for simple ways this might occur, we shall ask
that the limiting theory is not pathological.

An example of one such limit can be obtained by starting with the
assumption (true in supersymmetric theories, and in some sense in
Coleman-Weinberg scenarios of radiatively induced symmetry breaking)
that the quartic Higgs coupling $\lambda$ is proportional to $g^2$,
or $\alpha_{\rm gut}$:
\begin{equation}
\lambda \sim g^2 \sim \alpha_{\rm gut} \ .
\label{eq:1}\end{equation}
Then in order for the Higgs mass $\mu^2\sim \lambda v^2$ to remain
finite, we must have
\begin{equation}
v^2 \sim g^{-2} \sim \alpha^{-1}_{\rm gut} \ .
\label{eq:2}\end{equation}
If one demands that fermion masses remain finite and nonvanishing,
then the Yukawa couplings $h$ must be proportional to the gauge couplings $g$:
\begin{equation}
h^2 \sim g^2 \sim \alpha_{\rm gut} \ .
\label{eq:3}\end{equation}
Evidently, the gauge boson masses
\begin{equation}
m^2_{W,Z} \sim g^2v^2
\label{eq:4}\end{equation}
also remain finite and nonvanishing.

The net result in this ``second gaugeless limit" is a noninteracting
theory of {\em massive} quarks, leptons, gauge bosons, and Higgs bosons.
  Conceptually, it is a radical departure from the
conventional picture, if only because the Higgs Yukawa coupling
constants are proportional to gauge couplings.  How does this occur?  Is
it through symmetries, or dynamics, or geometry, or some combination?
Examples of this kind of behavior do exist in the literature, in terms
of attempts to relate the couplings through an assumed cancelation of
divergent radiative corrections between gauge and Higgs sectors
\cite{ref:e,ref:f}.

It is especially interesting that in this second gaugeless limit the
dependence of the standard model Lagrangian density on the coupling
constants $g \sim h$ is extremely simple.  After appropriate field
redefinitions, the residual dependence is an overall multiplicative
factor $1/g^2$ in front of the bosonic Lagrangian, with no dependence at
all within the fermionic sector (in the limit of only the top quark
possessing mass).  To see this, one simply rescales the gauge potentials
$A$ in the familiar way, and does the same with the Higgs fields
$\phi$ as well
\begin{equation}
g A \rightarrow A
\end{equation}
\begin{equation}
g \phi \rightarrow \phi \ .
\end{equation}
If one wishes, one may also rescale the fermion fields in the same way,
\begin{equation}
\psi \rightarrow g^{-1} \psi \,
\end{equation}
in which case the $g$-dependence of the entire Lagrangian density is
simply an overall coefficient $g^{-2}$.

\section{A Third Gaugeless Limit}

We now go still further and look for connections between the gauge/Higgs
parameters and the gravitational sector.  Our starting point is within
the gauge sector, and can be motivated by the hypothesis that gauge
bosons originate at the GUT scale as composites of other degrees of
freedom.  In more familiar contexts, this is expressed as a
compositeness condition \cite{ref:h},
\begin{equation}
Z_3 = \frac{g}{g_0} \rightarrow 0 \ ,
\label{eq:5}\end{equation}
where the limit $g_0 \rightarrow \infty$ implies compositeness: the
probability $Z_3$ of finding a bare boson within the physical boson
becomes zero in the limit.

The observed coupling $g$ is typically related to $g_0$ as follows
\begin{equation}
\frac{1}{g^2} = \frac{1}{g^2_0} + c \log \rightarrow c \log \ ,
\label{eq:6}\end{equation}
where the logarithm is typically an integral over contributions from
the boson's  constituents.  We now adopt the same structure, but using
gravitational parameters as arguments of the logarithm:
\begin{equation}
\frac{1}{\alpha_{\rm gut}} = \frac{4\pi}{g^2} \simeq
\frac{c_g}{4\pi}\, \ell n \frac{M^2}{\mu^2} \ .
\label{eq:7}\end{equation}
If the coefficient $c_g$ of the logarithm is chosen to be three, there
is good numerical agreement.  But no claim of a ``derivation" of that
coefficient, however, is implied, nor indeed of the functional form.

Once the gauge
couplings are expressed in such a way, it becomes reasonable to assume
that the Higgs couplings are also expressed in a similar way
\begin{eqnarray}
\frac{4\pi}{\lambda}
&\sim& \frac{c_\lambda}{4\pi}\, \ell n \, \frac{M^2}{\mu^2}
\nonumber \\[1ex]
\frac{4\pi}{h^2} &\sim& \frac{c_h}{4\pi}\, \ell n\, \frac{M^2}{\mu^2} .
\label{eq:8}\end{eqnarray}
Only the Higgs $vev$ remains to be estimated.  Given the dependence of the other
couplings upon the gravitational parameters, a natural choice is the rather
well-known relation \cite{ref:g}
\begin{equation}
v^2 \sim M\mu
\label{eq:9}\end{equation}
or, if one wishes  \begin{equation}
v^2 \sim M\mu\  \ell n\, \frac{M^2}{\mu^2} \ .
\label{eq:10}
\end{equation}
In this variant, the ``second gaugeless limit", is attained in the limit
\begin{eqnarray}
M &\rightarrow& \infty \nonumber \\
\mu &\rightarrow& 0 \\
0 &<& \hspace{-1ex} M\mu \ \   < \infty \ .
\label{eq:11}\end{eqnarray}
If one chooses to omit the logarithm in Eq.~(\ref{eq:10}), then one obtains
in the limit a noninteracting theory of {\em massless} quarks, leptons, Higgs
bosons, and gauge bosons.  Clearly both the massless and massive options should
be considered.

Numerically, one has for the value of the $v^2$
\begin{equation}
v^2 \sim6 \times 10^4 \  {\rm GeV}^2 \ .
\end{equation}
If one uses the Planck scale for $M$ in Eq. ({\ref{eq:9}}),
we obtain
\begin{equation}
M\mu \sim 10^7  \  {\rm GeV}^2
\label{eq:12}
\end{equation}
which is a little too large.  On the other hand, if the GUT scale of, say,
$3\times 10^{15}$
GeV is used in Eq.~({\ref{eq:10}}), then
\begin{equation}
M\mu \sim 10^3-10^4\  {\rm GeV}^2
\label{eq:13}
\end{equation}
which is a little too small.  Inclusion of unknown coefficients and/or
the logarithm can in principle provide the needed numerical agreement.
More important than that is to find even a hint of such behavior from an
underlying theory.

The above speculations can be expressed in differential form, in terms of
Gell-Mann-Low equations \cite{ref:i}.   Our basic premise is that the GUT
scale gauge and Higgs couplings are sensitive to the value of the cosmological
constant.  This sensitivity is to be expressed in terms of familiar-looking equations
\begin{eqnarray}
\mu\, \frac{dg^2}{d\mu} &=& \beta_g\, g^4 + \cdots \nonumber \\
\mu\, \frac{dh^2}{d\mu} &=&\beta_h\, h^4 + \cdots \\
\mu\, \frac{d\lambda}{d\mu} &=&\beta_\lambda \, \lambda^2 + \cdots \ .
\nonumber
\label{eq:14}
\end{eqnarray}
While these look like the usual renormalization-group equations, they are not.
They express the dependence of the usual running coupling constants, evaluated
at the GUT scale, upon the value of the cosmological constant.  While the general
form of the dependence has been assumed to be the same, these ``cosmological
$\beta$-functions" differ in detail; in particular the sign is changed for the
gauge couplings but not for the Higgs couplings.

We may also write a Gell-Mann-Low
equation for the Higgs $vev$.  Without the logarithm we have
\begin{equation}
\mu\, \frac{dv^2}{d\mu} = v^2 + \cdots
\label{eq:15}
\end{equation}
and with the logarithm
\begin{equation}
\mu \frac{dv^2}{d\mu} = v^2(1-\beta_g  g^2) + \cdots \ .
\label{eq:16}
\end{equation}

No matter which way this idea is expressed, the main question is whether
such a dependence of standard model parameters on the cosmological
constant is credible.  In its favor are the rough numerical agreements,
which we at least regard as unforced.  Also perhaps in favor of this
scenario is the feature that the dynamics becomes trivial, including the
vanishing of the $vev$, in the limit of vanishing of the cosmological
constant.  This is an avenue for at least reducing the electroweak
hierarchy problem to that of understanding the nature of the
cosmological constant.  And it clearly demands that the role of the
cosmological constant in the future theory be a central one.

\section {A Possible Connection to Cosmology}

Recently there has been a line of argument \cite{ref:aa} which utilizes
analogies of the standard model vacuum and its excitations with that of
quantum liquids, in particular with ${^3}$He-A.  In this visualization,
it is rather natural to expect a nearly vanishing vacuum pressure,
characteristic of an infinite liquid in equilibrium at zero temperature.
It is not much of a stretch to thereby obtain a vanishing vacuum energy
(cosmological constant) as well in that limit.  If the liquid has a
boundary, to be identified with an event horizon, then there will be
corrections, leading to a nonvanishing but small cosmological constant.
A rather concrete example of this general idea has been provided by the
picture of a black hole recently put forth by Chapline {\em et
al.}\cite{ref:bb} They assume that a phase transition occurs on the
horizon between the conventional exterior Schwarzschild black-hole
spacetime and an unconventional interior black-hole spacetime, taken to
be static de Sitter space.  This interior space possesses a cosmological
constant, which scales as follows:
\begin{equation}
\Lambda \sim \mu^4 \sim R^{-2}
\end{equation}
where $R$ is the radius of the black hole.  With the coupling constant
relations obtained in the previous section, this would imply that the
standard model parameters within the black hole differ from those
outside, in such a way that for infinite radius the gauge couplings and
particle masses vanish.  In the opposite limit of a Planck-radius black
hole, the gauge couplings become strong and the particle masses approach
the Planck scale.  If our universe contains a similar de Sitter horizon
\cite{ref:j}, then the standard model parameters will scale in a similar
way.  In particular, because of the above behavior of the cosmological
constant, the electroweak vacuum energy $v^4$ will scale as
\begin{equation}
v^4 \sim M^2\mu^2 \sim M^3R^{-1} \ .
\end{equation}
There will have to be a close connection between cosmology in the large,
in particular horizon structure, and the existence and nature of the
Higgs condensate \cite{ref:cc}.  This is reinforcement for the arguments regarding the
electroweak hierarchy problem made in the introductory section of this
note.

\section{Concluding Comments}

These ideas are of course extremely speculative.  Their value is in
proportion to what further, if anything, can be done with them.  We do
find encouragement in the numerics, and in the simplicity and cogency of
the relations which have been presented, especially with respect to the
Gell-Mann-Low equations for the couplings.  In the case of the
renormalization group equations of the standard model, the coefficients
are simple and calculable by essentially perturbative techniques.
Perhaps there is an analogously simple scheme (but radically different
in its physics!!) to be found.  And the fact that the logarithmic
factors, $\log M^2/ \mu^2$, associated with gauge and Higgs couplings only
appear as a multiplier of the entire standard-model Lagrangian density
might indicate that they are some kind of extra-dimensional phase
volume.
However, implementation of this idea in more concrete terms is beyond the
scope of this note.

\section{Acknowledgments}

I thank my colleagues at SLAC and Stanford University for useful
criticisms, and especially acknowledge most useful discussions with S.
Brodsky, H. Davoudiasl, C.~Taylor, G. Chapline, S. Dimopoulos, and R.
Laughlin.


\begin{thebibliography}{99}
\bibitem{ref:cc}
There cannot, however, be too strong a correlation between the horizon radius
and the scale size of the universe, as measured by the Hubble expansion, lest
the time dependence of standard-model couplings becomes unacceptably large.
\end{thebibliography}
\end{document}